# Analog Coupled Oscillator Based Weighted Ising Machine

Jeffrey Chou[1†*], Suraj Bramhavar[1†], Siddhartha Ghosh[1], William Herzog[1]

[1]*Massachusetts Institute of Technology Lincoln Laboratory, Lexington, Massachusetts, USA*

[†]Authors contributed equally
*jeff.chou@ll.mit.edu

**Abstract –** We report on an analog computing system with coupled non-linear oscillators which is capable of solving complex combinatorial optimization problems using the weighted Ising model. The circuit is composed of a fully-connected 4-node LC oscillator network with low-cost electronic components and compatible with traditional integrated circuit technologies. We present the theoretical modeling, experimental characterization, and statistical analysis our system, demonstrating single-run ground state accuracies of 98% on randomized MAX-CUT problem sets with binary weights and 84% with 5-bit weight resolutions. Solutions are obtained within 5 oscillator cycles, and the time-to-solution has been demonstrated to scale directly with oscillator frequency. We present scaling analysis which suggests that large coupled oscillator networks may be used to solve computationally intensive problems faster and more efficiently than conventional algorithms. The proof-of-concept system presented here provides the foundation for realizing such larger scale systems *using existing hardware technologies* and could pave the way towards an entirely novel computing paradigm.

1. Introduction

Solving certain classes of combinatorial optimization (CO) problems has proven to be notoriously difficult using standard von Neumann computing architectures. A canonical example is the traveling salesman problem, for which exact algorithms scale very poorly with problem size. Applications of CO problems span many disciplines, including business operations, scheduling, traffic routing, finance, big data, drug discovery, machine learning, and many other systems requiring the minimization of a complex energy landscape with multivariate inputs[1]. As the decades-long progress of digital CMOS technologies begins to plateau, there is a growing desire to find alternative computing methodologies which can address the challenges of these traditionally difficult problems. One strategy to solve CO problems lies in mapping them to the Ising Hamiltonian for spin glass systems and finding the ground state solution[2,3]. The Ising model dates back many decades but was re-popularized by D-Wave Systems in an attempt to exploit quantum mechanical phenomena to speed up computation. Novel quantum annealing machines [4–7] and digital CMOS annealing accelerators[8] are garnering significant attention in the hopes of solving such problems faster than conventional algorithms performing heuristic optimizations. The Ising Hamiltonian can also be mapped to the comparable field of Hopfield neural networks, which have been previously explored to solve CO problems[1,9–11].

Recently, alternative classical methods to solve the Ising model have emerged using optoelectronic parametric oscillators[12–14], memristor cross-bar arrays [11,15,16], electronic oscillators [17,18], and GPU based algorithms [19,20]. An analysis of one such coupled oscillator system revealed the potential for a significant speedup over digital computing algorithms at large node sizes[21]. Even algorithmic approaches which emulate nonlinear dynamical systems but are run on conventional computing hardware have been shown to match and even surpass the performance of state-of-the-art algorithms, motivating the desire to build these systems in physical hardware[22,23]. Scaling up the optoelectronic oscillator Ising machine remains challenging owing to its time-multiplexed architecture and highly complex and costly optoelectronic setup[12,13]. All of the proposed systems bear close resemblance to those described in the field of stochastic computing, where so-called 'p-bits', or probabilistic representations of digital bits have been shown in simulation to be capable of solving invertible logic and combinatorial optimization problems[3,24,25]. Analog implementations of these circuits typically rely on magnetic tunnel junctions, which could conceivably be implemented using standard CMOS technology, but a scalable physical implementation of this technology has yet to be realized. In a similar vein, the all-electronic oscillator concept initially proposed by Wang and Roychowdhury introduces the tantalizing prospect of creating a similar system using readily available electronic components interconnected in a parallel fashion and is particularly well suited for chip-scale integration and scaling using present day technologies.[17] In this case, the 'p-bit' is represented by the phase of an oscillator, which can probabilistically settle to one of two values. Previous reports have experimentally demonstrated a Chimera-graph architecture with a total of 240 nodes and described its performance on 20 random problems.[26] However, a rigorous and statistically significant analysis of the accuracy of this experimental system has not been demonstrated. In the reported physical implementation, resistive coupling is used to apply the connecting weights between their oscillators, where the absolute resistance values determines the coupling weights. However, specific details of the oscillator and interconnect architectures employed in these demonstrations are currently publicly unavailable, thus precluding our ability to determine the operation and limitations of this approach.

In this paper, we build upon this initial work and demonstrate a 4 node, fully-connected, differential LC (inductor-capacitor) oscillator based analog circuit with standard electronic components which accurately maps to the Ising model. One key difference between the circuit presented here and that demonstrated previously[17,26] involves the interconnection architecture and oscillator coupling scheme. Whereas previous implementations used a simple resistor to apply the connection weights directly, the circuit demonstrated in this work implements gain ratios in a cross-bar architecture. Specific advantages gleaned from using the cross-bar architecture include the ability to program weights with larger bit-depth and easy scalability in CMOS for large and fully-connected systems.[27] We provide a thorough analysis of system performance, describe in detail the circuit architecture including a new modality used to program variable interconnection strengths, and demonstrate the system capabilities in solving a variety of MAX-CUT problems with both binary and multi-bit weight values. To the best of our knowledge, this is the first demonstration of an all-electronic oscillator-based Ising machine with multi-bit weights. The thorough statistical analysis presented herein provides valuable insight into the viability of these systems as computing platforms when scaled to larger node counts.

## 2. Theory

Much of the theoretical framework for using coupled oscillators as Ising solvers has been described previously, but will be summarized here for convenience.[28] The Ising Hamiltonian is given by the following equation:

$$H = -\sum_{i,j}^{V} J_{ij} s_i s_j - \sum_{i}^{V} h_i s_i \qquad (1)$$

where $V$ represents the number of nodes in a particular problem set, $J_{ij}$ represents the weight values interconnecting the nodes, and $\boldsymbol{s} = [s_i \ldots s_V]$ represents the solution space where $s_i$ can take the value of either +1 (spin ↑) or -1 (spin ↓). One common benchmark optimization problem, known as the MAX-CUT problem, is defined by taking an undirected graph and finding a bisection of that graph which maximizes the cut set. This problem can be mapped directly to the Ising Hamiltonian above, and the solution to the problem is represented by the state $\boldsymbol{s}$ which minimizes $H$. A particular problem of interest is typically defined by a graph $G(V,E)$, where $V$ represents the number of vertices and $E$ represents the number of edges. An example of a 4-node system is shown in Fig. 1(a). If $J = 1$ for all connections, and we neglect the

Zeeman term ($h_i = 0$), the minimum solution is defined by 6 degenerate 2x2 solution sets, as shown in Fig. 1(b). It has been shown previously that these graphs can be represented by a network of coupled nonlinear oscillators whose phase dynamics are described by the Kuramoto model[29], and that this model maps directly to the Ising Hamiltonian if the phases of these oscillators take values of either 0° or 180°[17]. One mechanism used to polarize the phases is to introduce an injection-locking signal at twice the natural frequency of the oscillators[17,30]. This 'super-harmonic' injection locking signal, and its application in the context of LC-oscillator systems is mathematically similar to the case of the degenerate optical parametric oscillator used in previous optical Ising machines[12,13], where optical pump pulses at twice the optical oscillation frequency are used to create binary phase values.

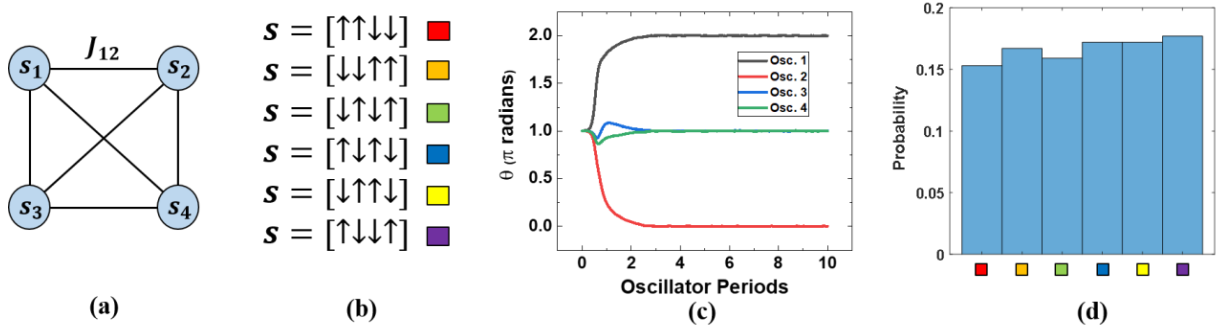

(a)  (b)  (c)  (d)

Fig. 1. (a) Schematic of a fully connected 4-node system. (b) Solution states of the degenerate 4 node state. (c) Simulation of the oscillator phases with initial conditions set to incorrect state. (d) Probability distribution of solution states after 1000 trials with randomized initial conditions.

In the LC-oscillator system, the phase evolution of each oscillator in the network ($\theta_i(t)$) can be represented by the following differential equation:

$$\frac{d\theta_i}{dt} = -\sum_{j=1}^{V} J_{ij} \sin(\theta_i(t) - \theta_j(t)) - A(t) \sin(2\theta_i(t)) \tag{2}$$

Where $V$ is the total number of oscillators (or vertices) and $A$ represents the amplitude of the injection-locking signal applied to each oscillator. The time $t$ in the equation is defined in dimensionless units in relation to the oscillation period. The addition of Gaussian phase fluctuations, representing noise in the system, converts equation Eq. (2) into a network of stochastic differential equations (SDEs), whose solutions can be approximated iteratively using the Euler-Maruyama method. Annealing is accomplished by gradually increasing the injection locking term $A$ according to the formula:

$$A(t) = A_0(1 - e^{-\frac{t}{\tau}}) \tag{3}$$

with $\tau$ set to 5 oscillation cycles. One solution simulated for the example problem described above is shown in Fig. 1(c), where the initial condition was intentionally selected to be an incorrect solution state ($s = [\uparrow\uparrow\uparrow\uparrow]$). Even in this scenario, the oscillator phases evolve to settle at one of the 6 correct solutions ($s = [\downarrow\downarrow\uparrow\uparrow]$). The system settles to the ground state within 3 oscillation cycles. This simulation was then run 1000 times with randomized initial conditions, and the system always settles to a correct solution state with fairly uniform probability (shown in Fig. 1(d)). The average settling time for the 1000 trials was 2.3 oscillation cycles.

### 3. Results
#### a. Circuit Implementation

The oscillator circuit employs a differential injection-locked frequency divider topology, as shown in Fig. 2(a)[31]. This architecture is selected to offer the option of synchronizing the oscillator with an incident super-harmonic injection locking signal. Transistors (Supertex TN0702) M1 and M2 form a cross-coupled pair, which serves as the negative resistance component, necessary for unity loop gain. The coupling signal from the other oscillators is applied differentially through transistors M3 and M4. This specific coupling circuitry is typically used for quadrature LC oscillators and employs an injection locking based coupling scheme, which has been previously mapped to the generalized Adler's equation[32], and can further be mapped to the Kuramoto model[28]. Current source I2 provide the bias current for the coupling signal. The output voltage of the oscillator is tapped at nodes VoL and VoR, directly out from the oscillating LC tank (L=100 µH, C=0.1 µF). The LC tank circuit has a resonant frequency of 50 kHz, and is composed of L1, C1, and L2,C2. Current source I1 provides a biasing current for the oscillator circuit and can also be used for injection locking to help polarize the phases to 0º and 180º.

A differential analog multiply and accumulate circuit in a cross-bar array is used apply the necessary signals into input nodes ViL and ViR, as shown in Fig. 2(b). The coupling coefficient polarity is controlled by the polarity of the output of the differential summing amplifier (Texas Instruments THS4140) to the nodes ViR and ViL. Digital potentiometers (Analog Devices AD5272), R12-R34, are employed to control the individual gains of each of the input oscillator signals. A key advantage of this interconnect circuit versus a more straight-forward resistive approach[17] stems from the ability to scale the number of fully-connected nodes without adding resistive loading to each oscillator by using a simple buffer circuit. The actual gain term is determined by the ratio of the feedback resistor, $R_{FB}$ = 1 kΩ, and the digital potentiometers. An Arduino microcontroller board was used to apply the digital I2C communication signals to the digital potentiometers. Precise tuning of the bias voltages for the oscillator and coupling circuit is critical to ensure accurate solution performance of the system.

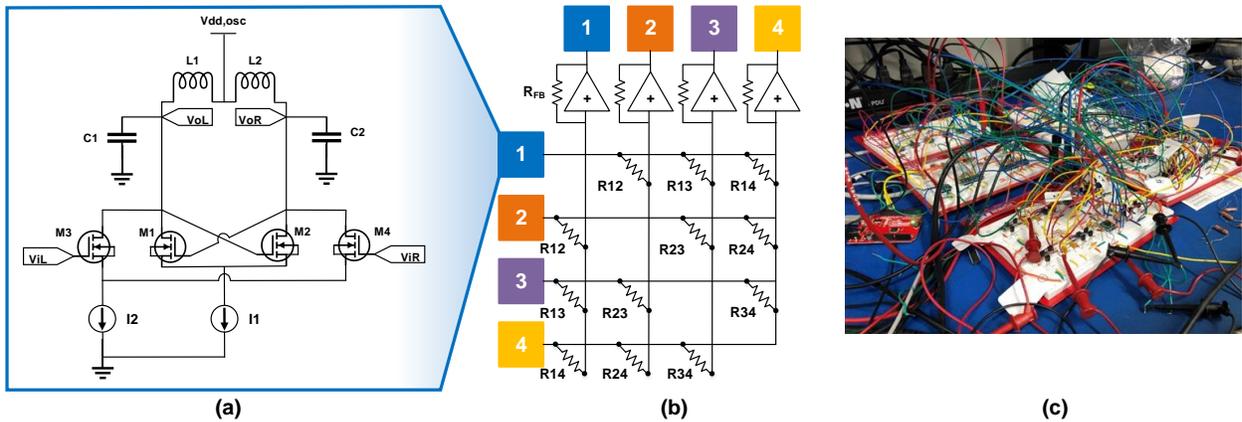

Fig. 2. (a) Circuit diagram of the LC oscillator circuit. The input coupling is inserted at the gates of transistors M3 and M4. (b) Multiply and accumulate cross-bar array composed of digital potentiometers. (c) Photograph of full breadboard system.

The analog coupling coefficients from the Ising Hamiltonian ($J_{ij}$) are mapped linearly to the ratio of the gains between the various oscillators. The digital potentiometers employed here have 1024 tap points, with a maximum resistance of 20 kΩ. The conversion from the analytical coupling coefficients to the digital potentiometers' programmed values ($D_{ij}$), which range from 1-1024, is shown the following equation:

$$D_{ij} = \frac{\beta\alpha}{J_{ij}} + \beta(1-\alpha) \qquad (4)$$

Where $\beta = R_{min} * 1024/20\text{k}\Omega$, and $\alpha$ is the mapping scalar. To prevent high currents at the summing amplifier at high coupling coefficient values, we set $R_{min} = 760\ \Omega$. The mapping from the coupling coefficients, $J_{ij}$, to the resistance values is scaled based on the $\alpha$ term to enable maximum dynamic range. A value of $\alpha = 2.5$ is used in this paper.

### b. Experimental Setup

A schematic of the experimental setup for the 4 node system is shown in Fig. 3(a). A National Instruments Labview program transmits the user-defined digital potentiometer values to an Arduino Uno board. The Arduino board then transmits the digital I2C resistors values to the six unique weights (J12 – J34). The analog oscillator voltage signals are monitored by the Labview program with a data acquisition board, at a maximum sampling rate of 500 kHz.

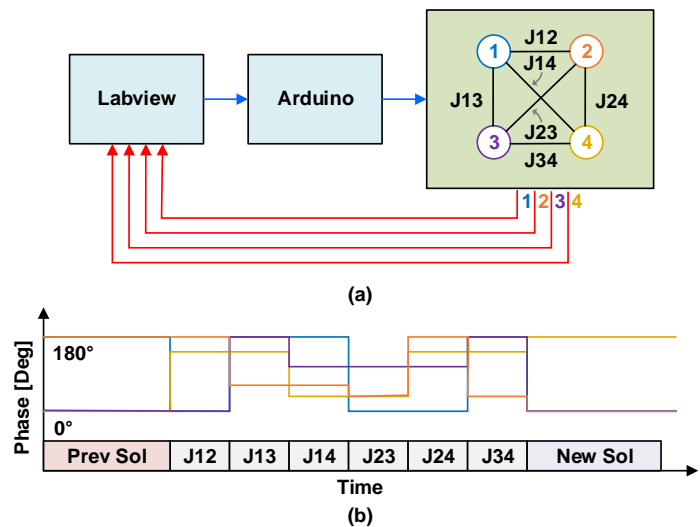

Fig. 3. (a) System level diagram of the oscillator control system. Blue lines indicate digital signals while red lines indicate analog signals. (b) Schematic timing diagram of the oscillator phases during system operation.

A representative phase-time trace of a single run is shown in Fig. 3(b). The digital potentiometers are programmed sequentially in time, due to the serial nature of the digital scheme. The time period of the weight programming is dependent on the loop speed of the Arduino board and the complexity of the code – which has a maximum loop frequency of 117 kHz.[33] After the last weight (J34) is programmed, the oscillators arrive at the new solution state. Phases of the oscillators are each measured relative to oscillator 1, whose phase is given a default binary value of "0". The binary phase outputs are then determined with a simple threshold, where phases differences > 90° with respect to oscillator 1 are assigned "1", and phases < 90º are assigned "0". We note that, unlike the simulation, no annealing of the coupling strength, is required to arrive at the solution. This is likely due to the low node count of the system. With more nodes, simulations of this oscillator network using coupled differential equations have shown that ramping up either the coupling strength or an external super-harmonic injection locking signal will be required in order for the system to preferentially settle to the correct solution. The super-harmonic injection locking signal serves a dual purpose in that it replaces the rounding feature described above, forcing the phases of each oscillator to their respective binary values (either 0º for "0" or 180º for "1"). As pointed out previously, this signal is required to extract sensible answers using large node count systems.[17]

A high sampling rate phase measurement of the phase transition of the oscillators is shown in Fig. 4. The phases were obtained via performing a moving window Fourier transform on the time-domain amplitude signal. The red dashed weight change line is when the last potentiometer (J34) has been digitally switched. From this measurement, we observe a 100 µs solution time, which translates to 5 cycles of the 50 kHz oscillators, as shown in Fig. 4(a). A potentially attractive pathway to scale the solution time of these systems involves simply increasing the oscillator frequency, as the settling time is expected to scale with oscillator period. In an attempt to validate this hypothesis, Fig. 4(b) shows the same experiment but with oscillator frequencies reduced by one order of magnitude, to 5 kHz, by utilizing 10 µF capacitors. We experimentally observe the time-to-solution also increases by approximately one order of magnitude, which suggests that the time-to-solution does indeed scale with the oscillator frequencies. The 90º threshold algorithm translates the final oscillator phase states to a binary solution of [0 1 1 0], if we use the order of the oscillator number. This solution is one of the 6 correct degenerate solutions for a fully-connected MAX-CUT problem with 4 nodes and equal connection weights.

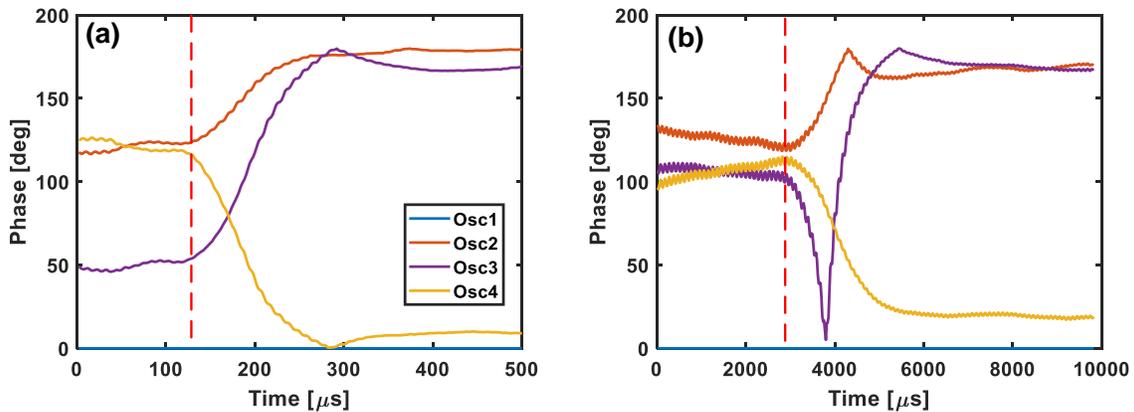

Fig. 4. Measured phase versus time data during the last weight change, J34, with (a) 50 kHz and (b) 5 kHz oscillators. A solution time of approximately 5 cycles is measured. The time-to-solution directly scales with oscillator frequency.

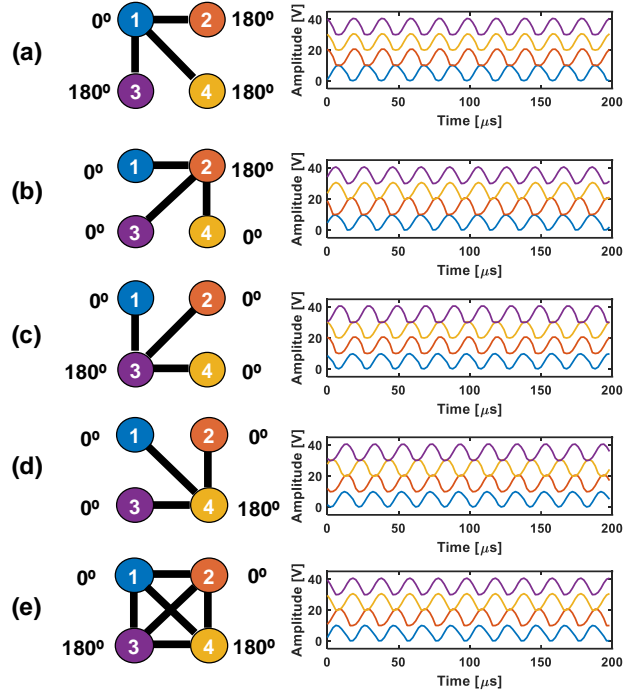

Fig. 5. (a)-(e) Measured oscillator amplitude data at various binary weight configurations. The lines connecting the oscillators correspond to a weight of 1, while an empty space corresponds to a weight of 0.

Time trace data of the oscillators under various binary weight configurations is shown in Fig. 5 to demonstrate the functioning system. The voltage amplitudes of each oscillator are shifted by 10 V to help distinguish the individual oscillators. All oscillation voltage amplitudes in this experiment range from 0 to 10 V. With the appropriate binary weights shown in Fig. 5(a)-(d), each oscillator can be made to be 180° out of phase with all other oscillators. With this graph structure of only one node connected to the remaining three, it intuitively follows that the energy minimization is achieved when the central node is out of phase with the others. Under the fully connected system in Fig. 5(e), there exist 6 degenerate solutions, in which any 2 oscillators are 180° out of phase with any other 2 oscillators. Here we only show one of the six solutions.

### c. Statistical Analysis

In order to validate the accuracy and reliability of the system, more than 2000 automated computer controlled experiments with random weights were performed. Fig. 6(a) and 6(b) shows the probability distribution of the arrival of each solution energy with a 1 bit and 5 bit weight values, respectively. The probability of achieving the ground state for 1 bit and 5 bit weights are 98% and 84%, respectively. The ground state probability was determined by 10 trials of each of the 2000 randomly generated problems. Clearly, the probability of achieving the ground state decreases with increasing weight bit resolutions. We observe that the solutions for all of the experiments follow a Boltzmann distribution with the peak of the curve at the ground state.

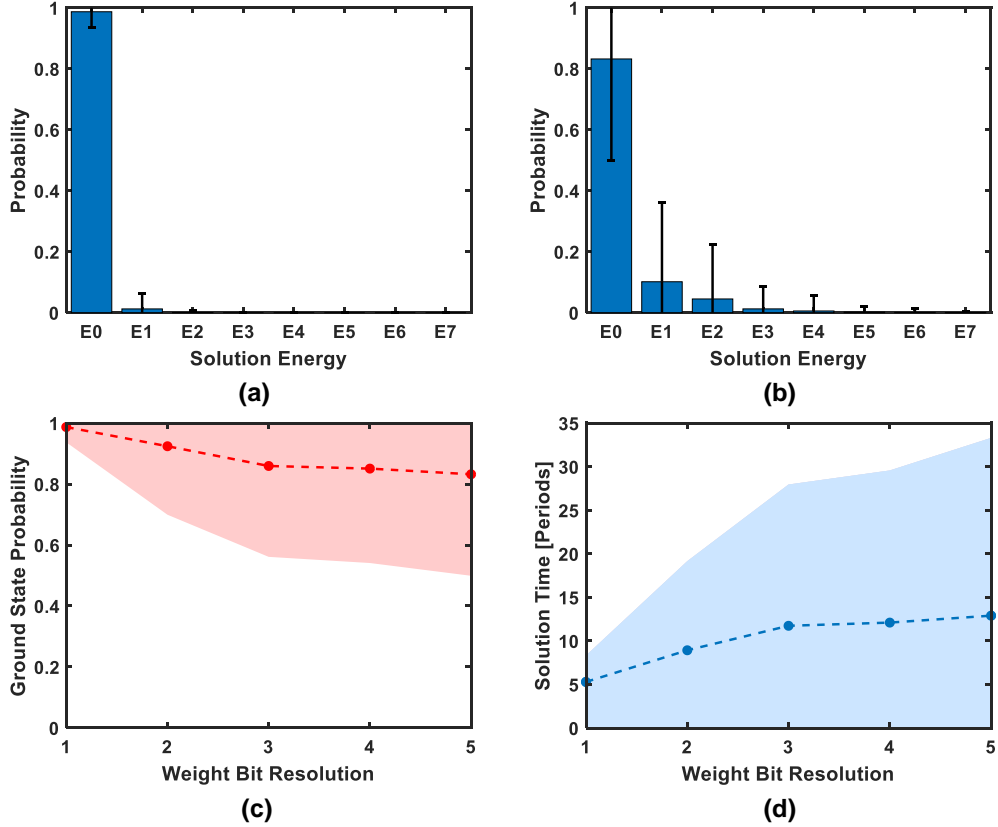

Fig. 6. Experimentally measured probability of solution energy for (a) 1 bit and (b) 5 bit weight resolutions after more than 2000 trials with random weights. Solution energy of E0 represents the ground state solution. Error bars represent the standard deviation. (c) The measured ground state probability versus weight bit resolution. (d) Estimated solution time in number of oscillator periods based on a 99% confidence. The corresponding shaded regions represent the standard deviation.

To fully characterize the ground state solution probability versus weight bit resolution, the results of running over 2000 random problems for weight bit resolutions from 1 to 5 is shown in Fig. 6(c). The ground state accuracy decreases monotonically as the bit resolution is increased. The total number of oscillator periods, $N_T$, required to reach ground state with 99% certainty is defined as $N_T = N_c \log(1 - 0.99) / \log(1 - P(b))$. Here, $P(b)$ is the measured ground state probability from a single run for bit resolution $b$, and $N_c$ is the number of oscillator periods for a single run. We obtain the $P(b)$ from Fig. 6(a) and plot the results in Fig. 6(d). Thus to achieve the ground state with 5 bits of weight resolution, an average of 13 periods would be required.

To characterize the cause of the non-ground state solutions for the non-binary weights, Fig. 7 plots the measured ground state probability versus the difference between the ground state Ising energy and the next highest energy state ($\Delta E = E1 - E0$) for randomly generated problems with 5 bit weights. Intuitively, this parameter characterizes the depth of the ground state potential well, with easier problems exhibiting large $\Delta E$, and more difficult problems exhibiting small $\Delta E$. The data shows that as $\Delta E$ decreases, the probability of finding the ground state on a single trial also decreases. For example, when $\Delta E < 2$, the probability of achieving the ground state falls below 0.9. Noise in the system causes difficulty in distinguishing energy-near solutions from the ground state solution. The minimum $\Delta E$ value inversely scales with weight bit resolution, which is the cause of the decreased accuracy for higher bit problems. Alternatively, binary weight systems typically have very high ground state probabilities due to the large $\Delta E$ values. For example, in the 4 node binary weight system, the minimum energy difference between E1 and E0, is $\Delta E = 2$, which gives rise to very high accuracy solutions. The complete absence of non-unitary probabilities for $\Delta E > 2$ implies that the system does not become trapped in local minimum, so long as the $\Delta E$ is large enough. To compare the experimental results to simulations, we overlay a similar plot generated by simulating a set of randomly generated 5-bit weight problems with the SDE simulation technique described above. The simulated sample set includes 1000 different problems, and each point represents the mean ground state probability value of all the problems with a given $\Delta E$ value. The behavior of the SDE simulated system matches reasonably well to the behavior measured experimentally.

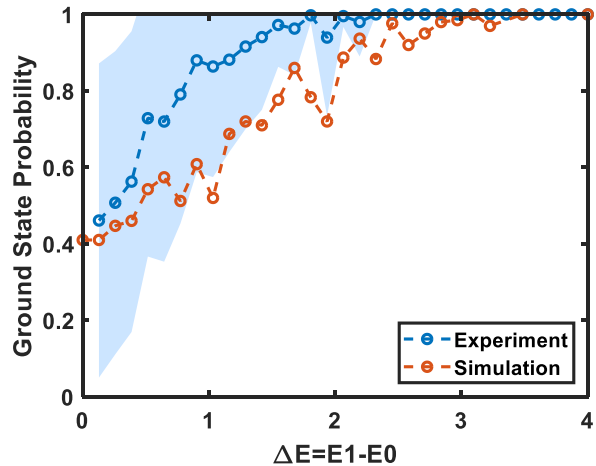

Fig. 7. Experimentally measured ground state probability versus solution energy difference between the ground state Ising energy and the next highest energy state ($\Delta E = E1 - E0$) for randomly generated 5 bit weights.

4. **Discussion**

To understand the expected performance as the system size is scaled, the SDE solver discussed previously was extended beyond the case of *V=4* and used to calculate the probability that the system can find the ground state as a function of *V*. We note that, for this scaling study, the annealing schedule described in Eq. (3) was used with $\tau = 5$ cycles. This value was intentionally chosen to be small to allow for a pessimistic assumption of the solution time scaling trends. The problem set for this test was a set of circular graphs, known as Mobius ladder graphs, with each node connected to its two nearest neighbors and one node directly opposite (shown in Fig. 8(a)). The ground-state solutions for this problem set were determined through brute force calculation[34], which allows for the solution from the SDE solver to be compared against a global energy minimum. Fig. 8(a) plots the probability *P* that the system reaches the ground state solution after 500 independent runs for each problem set, showing a natural roll-off from ~100% ground state probability when *V=8* to roughly 3% probability when *V=300*. Translating these results to calculate the overall time required to find a solution for a given problem ($t_s$), we first find $N_t$ by defining $N_c$ as the number of oscillation periods required for a single run of the simulated system to reach a minimum value and multiplying by the number of trials required to reach the ground state based on the expected probability *P* (as described in section 3c). Finally, multiplying $N_t$ by the period of one oscillation yields the final solution time $t_s$. Fig. 8(b) shows these computed values assuming a system built with 50 kHz oscillators (to match the current experimental circuit), along with a linear fit to a logarithmic trend line. The results are compared against the identical problem set and a similar analysis from a previous Ising machine implementations[13] (orange points). For illustrative purposes, a similar trend line is plotted for a system assuming an oscillator frequency of 1 MHz. Direct scaling with oscillator frequency was demonstrated in Fig. 4, and the additional line is shown here to highlight the potential speedup that can be achieved against existing implementations with relatively modest hardware specifications. It is also important to note that this scaling analysis focuses entirely on ground state solutions. As shown in Fig. 6(b), when the ground state solution is not reached, the system often settles to a value very close to the absolute minimum, which can be sufficient for many optimization problems.

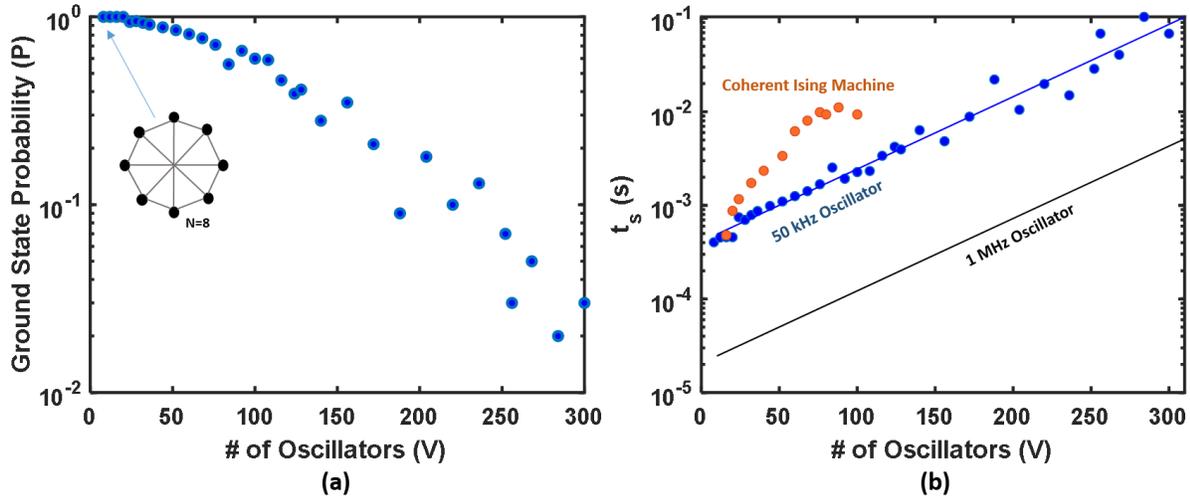

Fig. 8. (a) Simulated ground state probability on a circular graph versus number of oscillators. (b) Scaling simulation of the coupled oscillator system with 50 kHz and 1 MHz oscillator frequencies. For comparison, data extracted from previous Ising machine demonstrations for an identical problem[13] (orange dots) is plotted alongside the simulated data, illustrating a potential advantage of this coupled oscillator system.

Moving beyond simple Mobius ladder graph problems, we extend the analysis to randomized cubic graphs, where each node is connected to three others with the connections determined at random. For each node size, 100 different problem instances were generated and the exact ground state was found using the BiqMac MAX-CUT solver[35]. Each problem set was then run 100 times using the coupled oscillator simulator, and the ground state probability ($P$) and solution time ($t_s$) were determined as described previously. Fig. 9 shows the mean $t_s$ calculated for all 100 problem sets at each node size, assuming a 1 MHz oscillator. The data is plotted against an identical experiment on cubic graphs solved using a GPU-accelerated mean-field algorithm, which has shown ~20x speedup when compared against existing Ising machine implementations[19]. It is clear from the simulation that a crossover point is reached at *V ~ 150*, with the coupled oscillator approach trending much more favorably as the system size is scaled. As a point of reference, the BiqMac solver running on a modern CPU took approximately 10-20 seconds to solve random cubic problem instances with *V=200*. This analysis lends credibility to the promise of using coupled oscillator systems to solve non-trivial computational problems.

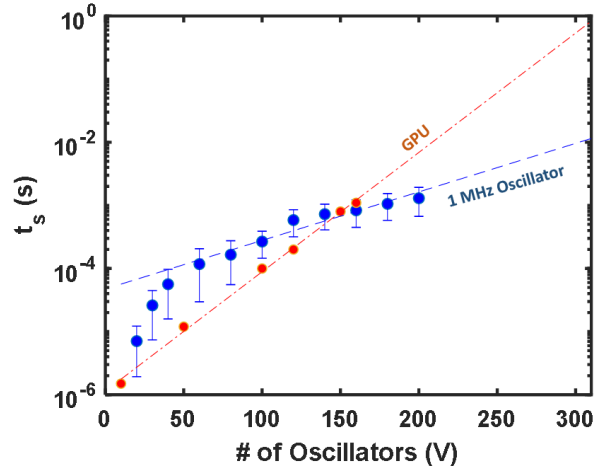

Fig. 9. Solution time scaling simulation for randomized cubic graphs using coupled oscillator approach (blue) vs. mean-field algorithm run on conventional GPU (orange). Dashed lines represent log-linear fits to the respective data. In the case of the coupled oscillator simulation, the fit was determined using the last 8 data points.

## 5. Conclusion

In summary we demonstrate the simulation, design, experiment, and characterization of a parallel, all-to-all connected coupled oscillator system which correctly maps to the weighted Ising model. The system is capable of successfully solving random MAX-CUT problems with 98% success probability with binary weights. The proof-of-concept breadboard experimental demonstration enabled a study of accuracy as a function of the interconnect weight bit-resolution. We show that the performance maps accurately to SDE simulations, and use these simulations to predict the system behavior at larger node sizes, which show favorable scaling when compared against existing optoelectronic implementations and GPU-based algorithms for similar problem sets. A full comparison on more densely connected problem sets will be the study of future work.

A significant advantage of the coupled oscillator system presented here lies in the fact that it can be scaled using existing low-cost hardware and lends itself especially well to an integrated circuit implementation. One major challenge, and an important area of future study, involves the interconnect architecture required to densely connect very large numbers of oscillators (>1e3). While time-multiplexed approaches have been implemented in recent instances[12,13], frequency- or code-division multiplexed systems could also provide interesting pathways to efficiently allocate connection resources. Another interesting area for study involves the use of measurement and feedback approaches to increase the ground state solution probability for larger problem sizes[36,37]. As transistor scaling begins to limit the progress in conventional digital computing architectures, the demand for alternative strategies continues to grow, and oscillator-based Ising machines could represent a novel and important platform to push the bounds of certain classes of computational problems.

## 6. References

1. Smith, K. A. Neural Networks for Combinatorial Optimization: A Review of More Than a Decade of Research. *Inf. J. Comput.* **11**, 15–34 (1999).

2. Lucas, A. Ising formulations of many NP problems. *Front. Phys.* **2**, (2014).

3. Yoshimura, C. *et al.* Uncertain behaviours of integrated circuits improve computational performance. *Sci. Rep.* **5**, 16213 (2015).

4. Johnson, M. W. *et al.* Quantum annealing with manufactured spins. *Nature* **473**, 194–198 (2011).

5. Nigg, S. E., Lörch, N. & Tiwari, R. P. Robust quantum optimizer with full connectivity. *Sci. Adv.* **3**, e1602273 (2017).

6. Puri, S., Andersen, C. K., Grimsmo, A. L. & Blais, A. Quantum annealing with all-to-all connected nonlinear oscillators. *Nat. Commun.* **8**, 15785 (2017).

7. Vinci, W. *et al.* Hearing the Shape of the Ising Model with a Programmable Superconducting-Flux Annealer. *Sci. Rep.* **4**, 5703 (2014).

8. Yamaoka, M. *et al.* A 20k-Spin Ising Chip to Solve Combinatorial Optimization Problems With CMOS Annealing. *IEEE J. Solid-State Circuits* **51**, 303–309 (2016).

9. Fang, Y., Yashin, V. V., Levitan, S. P. & Balazs, A. C. Pattern recognition with "materials that compute". *Sci. Adv.* **2**, e1601114 (2016).

10. Maffezzoni, P., Bahr, B., Zhang, Z. & Daniel, L. Oscillator Array Models for Associative Memory and Pattern Recognition. *IEEE Trans. Circuits Syst. Regul. Pap.* **62**, 1591–1598 (2015).

11. Guo, X. *et al.* Modeling and Experimental Demonstration of a Hopfield Network Analog-to-Digital Converter with Hybrid CMOS/Memristor Circuits. *Front. Neurosci.* **9**, (2015).

12. Inagaki, T. *et al.* A coherent Ising machine for 2000-node optimization problems. *Science* **354**, 603–606 (2016).

13. McMahon, P. L. *et al.* A fully programmable 100-spin coherent Ising machine with all-to-all connections. *Science* **354**, 614–617 (2016).

14. Hamerly, R. *et al.* Scaling advantages of all-to-all connectivity in physical annealers: the Coherent Ising Machine vs. D-Wave 2000Q. *ArXiv180505217 Phys. Physicsquant-Ph* (2018).

15. Shin, J. H., Jeong, Y. J., Zidan, M. A., Wang, Q. & Lu, W. D. Hardware Acceleration of Simulated Annealing of Spin Glass by RRAM Crossbar Array. in *2018 IEEE International Electron Devices Meeting (IEDM)* 3.3.1-3.3.4 (2018). doi:10.1109/IEDM.2018.8614698

16. Cai, F. *et al.* Harnessing Intrinsic Noise in Memristor Hopfield Neural Networks for Combinatorial Optimization. *ArXiv190311194 Cs* (2019).

17. Wang, T. & Roychowdhury, J. OIM: Oscillator-based Ising Machines for Solving Combinatorial Optimisation Problems. *ArXiv190307163 Cs* (2019).

18. Parihar, A., Shukla, N., Jerry, M., Datta, S. & Raychowdhury, A. Computing with dynamical systems based on insulator-metal-transition oscillators. *Nanophotonics* **6**, 601–611 (2017).

19. King, A. D., Bernoudy, W., King, J., Berkley, A. J. & Lanting, T. Emulating the coherent Ising machine with a mean-field algorithm. *ArXiv180608422 Quant-Ph* (2018).

20. Tiunov, E. S., Ulanov, A. E. & Lvovsky, A. I. Annealing by simulating the coherent Ising machine. *Opt. Express* **27**, 10288 (2019).

21. Haribara, Y., Utsunomiya, S. & Yamamoto, Y. A Coherent Ising Machine for MAX-CUT Problems: Performance Evaluation against Semidefinite Programming and Simulated Annealing. in *Principles and Methods of Quantum Information Technologies* (eds. Yamamoto, Y. & Semba, K.) 251–262 (Springer Japan, 2016). doi:10.1007/978-4-431-55756-2_12

22. Di Ventra, M. & Traversa, F. L. Perspective: Memcomputing: Leveraging memory and physics to compute efficiently. *J. Appl. Phys.* **123**, 180901 (2018).

23. Di Ventra, M. & Pershin, Y. V. The parallel approach. *Nat. Phys.* **9**, 200–202 (2013).

24. Sutton, B., Camsari, K. Y., Behin-Aein, B. & Datta, S. Intrinsic optimization using stochastic nanomagnets. *Sci. Rep.* **7**, 44370 (2017).

25. Camsari, K. Y., Faria, R., Sutton, B. M. & Datta, S. Stochastic $p$-Bits for Invertible Logic. *Phys. Rev. X* **7**, 031014 (2017).

26. Wang, T., Wu, L. & Roychowdhury, J. Late Breaking Results: New Computational Results and Hardware Prototypes for Oscillator-based Ising Machines. *ArXiv190410211 Cs* (2019).

27. Analogue signal and image processing with large memristor crossbars | Nature Electronics. Available at: https://www.nature.com/articles/s41928-017-0002-z. (Accessed: 5th August 2019)

28. Wang, T. & Roychowdhury, J. Oscillator-based Ising Machine. *ArXiv170908102 Phys.* (2017).

29. Acebrón, J. A., Bonilla, L. L., Pérez Vicente, C. J., Ritort, F. & Spigler, R. The Kuramoto model: A simple paradigm for synchronization phenomena. *Rev. Mod. Phys.* **77**, 137–185 (2005).

30. Wang, T. & Roychowdhury, J. PHLOGON: PHase-based LOGic using Oscillatory Nano-systems. in *Unconventional Computation and Natural Computation* (eds. Ibarra, O. H., Kari, L. & Kopecki, S.) 353–366 (Springer International Publishing, 2014).

31. Rategh, H. R. & Lee, T. H. Superharmonic injection-locked frequency dividers. *IEEE J. Solid-State Circuits* **34**, 813–821 (1999).

32. Mirzaei, A., Heidari, M. E., Bagheri, R., Chehrazi, S. & Abidi, A. A. The Quadrature LC Oscillator: A Complete Portrait Based on Injection Locking. *IEEE J. Solid-State Circuits* **42**, 1916–1932 (2007).

33. Go Speed Racer...Arduino Speed Test - learn.sparkfun.com. Available at: https://learn.sparkfun.com/blog/1687. (Accessed: 18th May 2019)

34. Godsil, C. & Royle, G. F. *Algebraic Graph Theory*. (Springer-Verlag, 2001).

35. Rendl, F., Rinaldi, G. & Wiegele, A. Solving Max-Cut to optimality by intersecting semidefinite and polyhedral relaxations. *Math. Program.* **121**, 307 (2008).

36. Leleu, T., Yamamoto, Y., McMahon, P. L. & Aihara, K. Destabilization of Local Minima in Analog Spin Systems by Correction of Amplitude Heterogeneity. *Phys. Rev. Lett.* **122**, 040607 (2019).

37. Albash, T., Martin-Mayor, V. & Hen, I. Analog errors in Ising machines. *Quantum Sci. Technol.* **4**, 02LT03 (2019).

### 7. Author Contributions Statement

## 8. Competing Financial Interests

The authors declare the following patent application: U.S. Patent Application No.: 62/826,080.

## 9. Additional Information

DISTRIBUTION STATEMENT A. Approved for public release. Distribution is unlimited.

This material is based upon work supported by the United States Air Force under Air Force Contract No. FA8702-15-D-0001. Any opinions, findings, conclusions or recommendations expressed in this material are those of the author(s) and do not necessarily reflect the views of the United States Air Force.